\documentclass[aip,graphicx]{revtex4-2}  
\usepackage[english]{babel}

\draft 

\usepackage{graphicx}
\usepackage{dcolumn}
\usepackage{bm}

\usepackage[utf8]{inputenc}
\usepackage[T1]{fontenc}
\usepackage{mathptmx}
\usepackage{etoolbox}
\usepackage{plimsoll}
\usepackage{soul} 
\usepackage{mathtools}
\usepackage{miller}
\usepackage{color}

\makeatletter
\def\@email#1#2{%
 \endgroup
 \patchcmd{\titleblock@produce}
  {\frontmatter@RRAPformat}
  {\frontmatter@RRAPformat{\produce@RRAP{*#1\href{mailto:#2}{#2}}}\frontmatter@RRAPformat}
  {}{}
}%
\makeatother
\begin{document}


\title{Pseudo-grand canonical molecular dynamics \emph{via} volumetrically controlled osmotic pressure} 



\author{Blake I. Armstrong}
\author{Aaron D. Copeland}
\affiliation{ 
School of Molecular and Life Sciences, Perth, Western Australia, Australia
}
\author{Davide Donadio}
\affiliation{%
Department of Chemistry, University of California Davis, Davis, USA
}%
\author{Paolo Raiteri}
\email{p.raiteri@curtin.edu.au}
\affiliation{ 
School of Molecular and Life Sciences, Perth, Western Australia, Australia
}


\date{\today}

\begin{abstract}
Molecular dynamics simulations are typically constrained to have a fixed number of particles, which limits our capability to simulate chemical and physical processes where the composition of the system changes during the simulation time. 
Typical examples are the calculation of nucleation and crystal growth rates in heterogeneous solutions where the driving force depends on the composition of the fluid.
Constant chemical potential molecular dynamics simulations would instead be required to compute time-independent growth and nucleation rates. 
While this can, in principle, be achieved through the addition and deletion of particles using the grand canonical partition function, this is very inefficient in the condensed phase due to the low acceptance probability of these events. 
Adaptive resolution schemes, which use a reservoir of non-interacting particles that can be transformed into solute particles, circumvent this problem, but at the cost of relatively complicated code implementations. 
In this work, a simpler approach is proposed that uses harmonic volumetric restraints to control the solute osmotic pressure, which can be considered a proxy for the system's chemical potential. 
The osmotic pressure regulator is demonstrated to reproduce the expected properties of ideal gases and ideal solutions. 
Using the mW water model, the osmotic pressure regulator is shown to provide a constant growth rate for ice in the presence of an electrolyte solution, unlike what standard molecular dynamics simulations would produce.
\end{abstract}


\maketitle 


\section{Introduction}

Molecular dynamics (MD) simulations are fundamentally important in computational chemistry, providing atomic-level insights into complex physical and chemical processes. Among various MD techniques, grand canonical molecular dynamics (GCMD) is a powerful approach for studying systems that are in thermodynamic equilibrium with an external reservoir such that the total chemical potential ($\mu$) of the system is maintained constant by exchanging particles and energy with the reservoir.
The theoretical foundation for constant chemical potential simulations emerged from statistical mechanics work performed by Norman and Filinov\cite{filinov1969TVT} as well as Adams\cite{adams1974MP} using Monte Carlo methods to simulate fluids.  
Simulations in the $\mu$VT ensemble (constant $\mu$, volume, and temperature) are based on the Widom particle insertion idea,\cite{widom_topics_1963} which connects the Boltzmann factor associated with the insertion of a particle in a fluid to the excess chemical potential. The constant $\mu$ is obtained by accepting or rejecting insertion and deletion moves.
These methods saw continued development alongside improving computational capabilities\cite{rowley1975JCP,yao1982MP,torrie1977JCP,yau1994JCP}, offering new possibilities for studying phenomena where particle exchange with a reservoir is essential and traditional microcanonical (NVE), canonical (NVT), and isothermal-isobaric (NPT) ensembles are inadequate. 
Inevitably, grand canonical methods were extended to MD simulations under various formalisms\cite{cagin1991MS,cagin1991MP,weerasinghe1994MP,ji1992JCP,lynch1997JCP,lo1995JCP,palmer1994JCP,shroll1999JCP,heffelfinger1994JCP,eslami2007JCP}, which proved valuable in investigating systems where traditional MD approaches fall short, including studies on adsorption phenomena in nanoporous materials\cite{hoang2012JCP, dubbeldam2007MS}, membrane transport processes\cite{mollahosseini2021JMGM,zhai2021CJCE}, and interfacial systems\cite{karmakar2023ACR}, where particle exchange results in a significant change in the chemical potential of the system. Particularly relevant are interfacial solution processes, which are driven by solute activity. Crystal growth and nucleation processes draw matter from solution to be incorporated into a solid crystallite or fluid-like agglomeration of particles. In closed systems, the growth of a crystalline nucleus, or the progression of a crystallising front, depletes the finite reservoir of solute particles, thereby modifying the solute's chemical potential and reducing the driving force towards crystallisation.\cite{sosso_7sins} 

Despite its usefulness, the implementation and application of GCMD have presented numerous challenges that continue to drive method development in the field. 
The insertion and deletion of particles create discontinuities in the potential energy surface, requiring careful consideration of system stability and accurate integration of the equations of motion.\cite{boinepalli2003JCP,cagin1991MP} 
Selecting appropriate insertion sites, particularly in dense systems, remains computationally challenging and can significantly impact simulation efficiency.\cite{attard1997JCP, shi2001FPE} Properly handling long-range interactions and maintaining temperature control during particle exchange events continue to pose technical challenges.\cite{boinepalli2003JCP} The dynamic nature of particle numbers in GCMD also presents unique challenges for parallel efficiency in MD engines.\cite{guo2018JTCC}

A natural consequence of the difficulties described above is the development of techniques that circumvent the explicit deletion/insertion of particles. A particularly noteworthy technique in this fashion is the AdResS algorithm,\cite{praprotnik_adaptive_2005,wang2013PRX} which represents an external particle reservoir at a lower, coarse-grained resolution compared to the full atomistic resolution of the main simulation section. Particles can then exchange between the two regions \emph{via} a hybrid region, whereby their resolution is adaptively changed depending on the flow direction. The AdResS technique has been subsequently expanded upon by reformulation in terms of a global Hamiltonian (H-AdResS)\cite{potestio2013PRL,potestio2013PRL2}, which improves upon the thermodynamic control of the atomistic and coarse-grained regions, with continued development in proper treatment of long-range interactions.\cite{heidari2016EPJST} AdResS and H-AdResS have been successfully applied to various systems, including water simulations, methanol-water mixtures, and triglycine in aqueous urea.\cite{mukherji_polymer_2014,potestio2014E,heidari_steering_2019,heidari_open-boundary_2020} 
Importantly, these methods reduce the number of force calculations in coarse-grained regions while maintaining thermodynamic properties between decoupled regions. In the context of crystal growth simulations, this effectively increases the number of solute particles from which the atomistic region can draw to maintain constant chemical potential. Care must be taken when compensating for thermodynamic imbalances between regions of different resolutions, especially when accounting for finite-size effects in free energy calculations.
However, the implementation of adaptive multiscale methods in well-established software packages is complicated, as it involves deep changes in the core functions of the codes and a difficult implementation of the load balance for parallel computing, effectively hampering computational performance.\cite{heidari2016EPJST} 

Another important technique that circumvents the explicit insertion/deletion of particles is the C$\mu$MD method, which effectively maintains a constant chemical potential of an explicitly defined `control region' within a solution by applying an external force to the solute to maintain a desired concentration.\cite{perego2015JCP} This technique has been applied to study nucleation and self-assembly, adsorption processes, and membrane permeation and separation,\cite{karmakar2023ACR}
maintaining constant solute concentrations during crystal growth and nucleation processes. 
Similar techniques have been proposed for solution processes that regulate the chemical potential \emph{via} osmotic pressure. For instance, phase separation and coexistence have been simulated using the `osmotic Gibbs-ensemble' technique whereby a semi-permeable membrane interfaces two distinct phases.\cite{brennan2002M,banaszak2004JCP} Solvent transport between the membrane maintains constant osmotic pressure (and consequently the chemical potential) of an idealized polymer solute, thus removing the need for explicit insertion/deletion of polymer molecules. This technique has been applied to study the crystal nucleation of sulfamerazine from acetonitrile/methanol by maintaining an effective solute supersaturation.\cite{liu2018MP} 

Building upon these foundational approaches, we combine the idea of computing the osmotic pressure using a harmonic restraining potential\cite{luo_roux,yoshida2017JCP} alongside the ideas in C$\mu$MD, which apply forces to solute particles to control the chemical potential of solutes within defined volumes. 
To this aim, we have developed a Berendsen-like pseudo-GCMD algorithm that maintains a constant osmotic pressure in a portion of the simulation cell by using an impermeable membrane acting on the solute particles, which can be seen as a proxy for constant chemical potential.

According to the foundational work by Lewis,\cite{Lewis1908JACS} the general form of osmotic pressure is given by;
\begin{equation}\label{eq:osmotic_1}
    \Pi - \frac{1}{2}\kappa_T\Pi^2 = - \frac{RT}{V_0}\ln(1-\chi)
\end{equation}
where $\Pi$ is the osmotic pressure, $\kappa_T$ is the compressibility of the pure solvent, $V_0$ is the molar volume of the solvent and $\chi$ is the molar fraction of the solute
\begin{equation}
    \chi = \frac{n_{solute}}{n_{solvent}+n_{solute}}.
\end{equation}
At moderate molar fractions, the dependence of the osmotic pressure on the solvent compressibility is negligible, allowing the expression to be simplified to 
\begin{equation}\label{eq:osmotic_2}
    \Pi = - \frac{RT}{V_0}\ln(1-\chi).
\end{equation}
where higher-order terms of $\chi$ have been neglected.
Assuming dilute solutions and incompressible fluids, a series of approximations can be made:
\begin{itemize}
    \item $\ln(1-\chi) \approx -\chi$
    \item $\chi \approx n_{solute}/n_{solvent}$
    \item $V_0 \approx V_{solution} / n_{solvent}$
\end{itemize}
\noindent which reduces Eq.~\ref{eq:osmotic_2} to the well-known van't Hoff equation for the osmotic pressure of dilute solutions
\begin{eqnarray}\label{eq:osmotic_3}
    \Pi &\approx& RT\ c
\end{eqnarray}
\noindent where $c=n_{solute}/V$ is the molar concentration of the solute.

Given the fundamental relationships between chemical potential $\mu$ and thermodynamic activity ($a$), and mole fraction $\chi$, 
\begin{equation}\label{eq:act_mu}
\mu_{i} = \mu_{i}^{\stst} + RT\ln{a_{i}}
\end{equation}
\begin{equation}\label{eq:act_mf}
a_i = \gamma_i \chi_i
\end{equation}
where $\mu_{i}^{\stst}$ is the standard chemical potential and $\gamma_i$ is the activity coefficient of species $i$, we can obtain an important thermodynamic insight. 
In fact, these equations reveal a critical thermodynamic principle (given constant temperature and external pressure), that when the osmotic pressure $\Pi$ is maintained constant (as in Eq.~\ref{eq:osmotic_1}), the mole fraction $\chi$ remains invariant. Consequently, the thermodynamic activity $a$ remains unchanged ( Eq.~\ref{eq:act_mf}). 
This, in turn, ensures that the chemical potential $\mu$ stays constant, as directly shown by Eq.~\ref{eq:act_mu}. In fact, given a finite number of solute particles in a closed system, the chemical potential can be effectively maintained without explicit computation by maintaining a constant, activity-dependent, external osmotic pressure acting on the solute. 

Herein, we present the implementation of an osmotic regulator as a custom external force within the OpenMM\cite{openmm} MD simulation engine and validate its functionality \emph{via} three distinct approaches.
First, the osmotic regulator is demonstrated to effectively regulate the pressure and volume of argon gas to align with the predictions of the ideal gas law; and the fluctuations in the pressure and volume of the osmotic membrane are compared with those obtained from standard MD simulations using the Berendsen\cite{berendsen1984JCP} and Monte-Carlo barostats. 
Second, the osmotic pressure regulator is used to quantitatively reproduce the predictions made by Equation \ref{eq:osmotic_1} for an ideal solution containing various solute molar fractions. 
Lastly, the growth rate of ice under constant osmotic pressure conditions is investigated.
By comparing scenarios where ice is put in contact with neat water or electrolyte solution, with and without the osmotic pressure regulator, we demonstrate how the algorithm proposed here is capable of producing a constant growth rate, even when a fraction of the solute is incorporated in the growing crystal, unlike standard MD simulations.

These points clearly demonstrate the accuracy and practical utility of the algorithm used to regulate the osmotic pressure. Importantly, due to OpenMM's purposefully designed, highly customisable nature, incorporating the osmotic regulator has minimal impact on the simulation's performance time (given reasonable input parameters). The effects of various parameters on the fluctuations of the osmotic pressure and membrane volume are also investigated to highlight a range of sensible values of the parameters used in the algorithm. 

\section{Methods}

Luo and Roux introduced a simple method to compute the osmotic pressure of an electrolyte solution directly from MD simulation by using a harmonic restraint to mimic the presence of a semipermeable membrane.
A similar approach was also reported by Bocquet and co-workers, who used a 1-dimensional restraining potential to compute the non-equilibrium diffusio-osmotic flow from non-equilibrium molecular dynamics simulations.\cite{yoshida2017JCP,marbach2019CSR}
Following their footsteps, we compute the instantaneous osmotic pressure as the sum of the forces acting on the solute particles due to the osmotic restraint ($\vec{\nabla} U_i(r)$) divided by the surface area of the restraining membrane.\cite{luo_roux,yoshida2017JCP} 
In order to account for the direction of the restraint and for non-planar restraining geometries, we can simply project the restraining force onto the direction normal to the membrane ($\vec{n}$)
\begin{equation}
\Pi = \frac{1}{A}\sum_{i\in solute} -\vec{\nabla} U_i(r) \cdot \vec{n}.
\end{equation}
For planar and slab geometries, the surface area is simply calculated as the area of the simulation cell parallel to the restraint (times two for the slab geometry). 
In contrast, the surface areas for the cylindrical and spherical restraints are not as straightforward, as their radii $R$ do not represent infinitely hard walls but instead is a soft potential that allows solute particles to diffuse beyond $R$.
Therefore, the surface area of the spherical and cylindrical restraints needs to be computed from the derivative of the effective volume of the restraints, which depend on the magnitude of $K$, $R$ and the system's temperature $T$. 
The full integral derivation of these quantities is shown in the appendix.

\subsection{Osmotic regulator}
\begin{figure}[htbp]
    \centering
    \includegraphics[width=0.75\columnwidth]{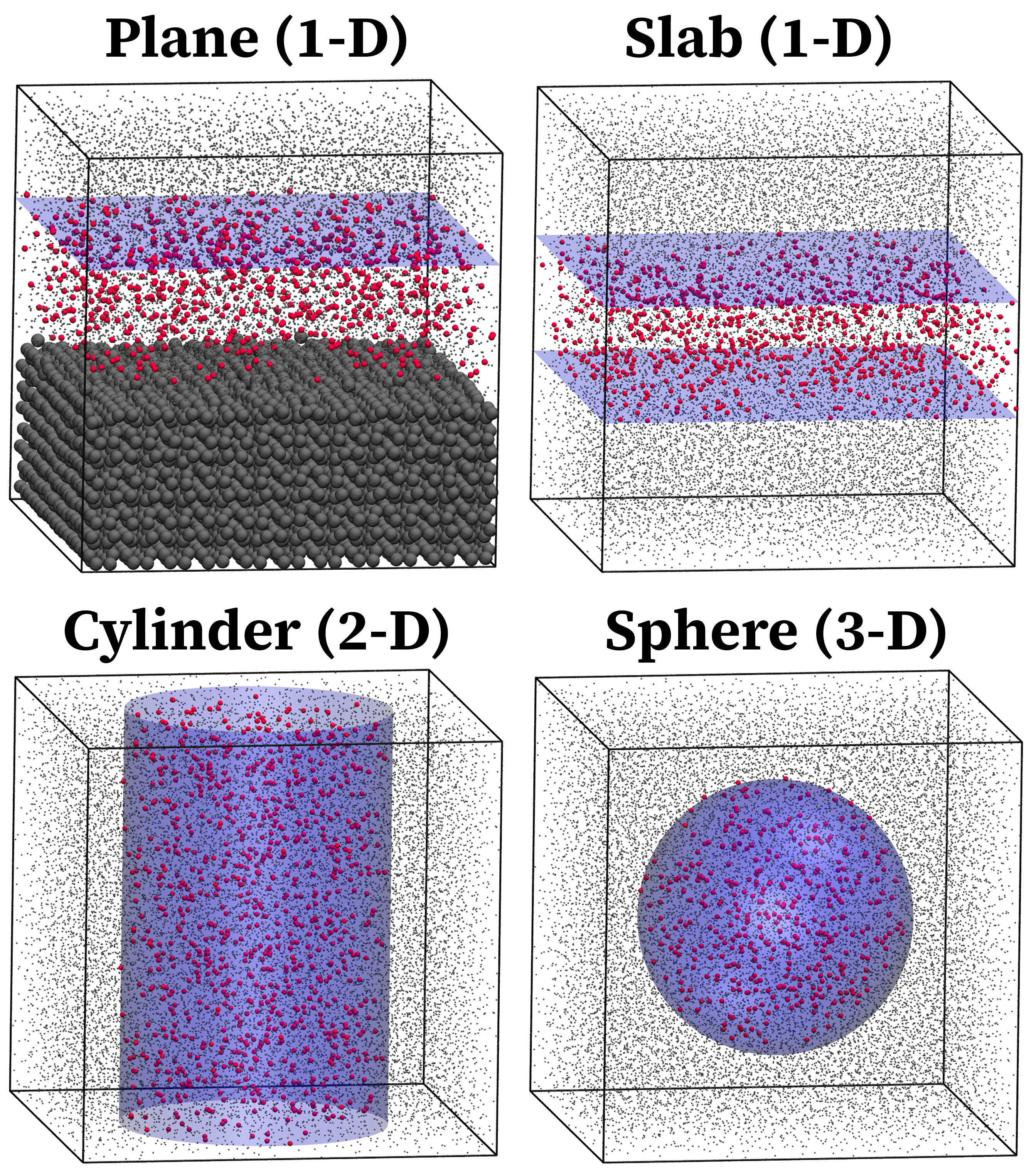}
    \caption{Graphical representations of the four available geometry types of the osmotic membrane. Each cube, outlined in solid black lines, denotes the bounds for the unit cell used to run MD. The small black points/dots depict the location of a solvent; the larger grey spheres in the plane geometry unit cell represent particles within a solid/crystallite; the red spheres depict solute particles that the osmotic membrane acts on; and the transparent blue surfaces depict the start of the harmonic potential (outside of the flat-bottomed region). }
    \label{fig:geometries}
\end{figure}

The osmotic membrane is constructed as a flat-bottomed harmonic potential,
\begin{equation}\label{eq:flat_bottom}
U(r) = 
\begin{dcases*}
    0 & if  $\lvert r\rvert \leq$ R \\
    \frac{1}{2}K(r-R)^2                   & \text{otherwise}
\end{dcases*}
\end{equation}
\noindent where $R$ is the length of the flat-bottom portion of the potential, $K$ is the spring constant, and $r$ is the distance of the particle from an arbitrary centre point that defines the geometry of the potential. $U(r)$ is applied to a selection of particles defined as the solute to which the external osmotic pressure should be applied. In this way, $U(r)$ effectively constrains the solute particles to a volume dependent on the dimensionality of $r$ and the magnitude of $R$. The various geometries that result from changing the dimensionality of $r$ are described as such; 
\begin{align*}
\textbf{Plane:} & \quad
r = x-x_0 \\
\textbf{Slab:} & \quad
r = \sqrt{(x-x_0)^2} \\ 
\textbf{Cylinder:} & \quad
r = \sqrt{(x-x_0)^2 + (y-y_0)^2} \\ 
\textbf{Sphere:} & \quad r = \sqrt{(x-x_0)^2 + (y-y_0)^2 + (z-z_0)^2}
\end{align*}

\noindent where $x_0$, $y_0$ and $z_0$ denote the user-defined locations for the centre point of the corresponding geometry, \emph{e.g.}, defining $r = \sqrt{(x-x_0)^2 + (y-y_0)^2}$ for the cylinder geometry creates a cylinder perpendicular to the $z$ axis centred about $(x_0,y_0)$ with a radius $R$. 
The orientation of the cylinder, slab and plane can be easily modified by swapping $x$ and $y$ with other Cartesian directions.
The corresponding graphical representations for each geometry are shown in Figure \ref{fig:geometries}. 
While multiple geometries can be added on top of each other simultaneously to create complicated geometrical restraints, these four shapes are already sufficient for most applications in MD simulations.
In particular, the slab, cylinder and spherical geometries appear particularly useful to study nucleation of salts, while the plane geometry would be more suited to study the surface growth kinetics, where only one restraining plane is required. 
It is worth mentioning that in this case, the bottom of the solid slab needs to be restrained to prevent translation of the entire system during the MD simulation.

The osmotic force regulator described here follows a Berendsen-like formalism\cite{berendsen1984JCP} where there is a target osmotic pressure $\Pi_0$, an ensemble osmotic pressure $\Pi_{ens}$, a relaxation time $\tau$, and an isothermal compressibility, $\kappa_T$, which dampens the fluctuations in the osmotic pressure;
\begin{equation}\label{eq:berendsen}
    \lambda = 1 - \frac{\kappa_T \Delta T}{\tau} (\Pi_0 - \Pi_{ens})
\end{equation}
where $\lambda$ describes the scaling factor needed to resize the volume of the osmotic membrane. 
Notably, the formalism described in Equation~\ref{eq:berendsen} does not use the instantaneous pressure, but rather $\Pi_{ens}$, which is a rolling average of the instantaneous osmotic pressure sampled every $N$ steps, where $N$ is referred to as the `compute-interval' (CI). 
The length of the rolling average is dubbed the `sample length' (SL) in the remainder of this work.
This is important, as the algorithm described here does not rescale the particle's positions in the same way as the Berendsen barostat, so using the instantaneous osmotic pressure would result in massive fluctuations in the position of the membrane, which would make the simulation unstable. 
The scaling parameter $\lambda$ is then used to modify the position of the restraining potential ($R$), driving the osmotic pressure towards $\Pi_0$, by changing the restraint volume
\begin{equation}
R(t+\Delta t) = \lambda^{1/d} R(t)
\end{equation}
where $d$ is the dimensionality of the restraining membrane; $d=1$ for plane and slab geometries, $d=2$ and $d=3$ for cylindrical and spherical membranes.
Unless otherwise stated, a value of 0.01~bar$^{-1}$ was used for $\kappa_T$, which is demonstrated later to produce reasonable fluctuations in the osmotic pressure with sensible values of $\tau$. While the main objective of the osmotic pressure membrane is to maintain an externally applied osmotic pressure, it can also simply record the osmotic pressure/volume without rescaling $R$, which could be useful to compute the solute activity coefficient.

\subsection{Molecular dynamics}

The implementation of the osmotic pressure regulator algorithm in OpenMM was first validated by demonstrating that it can be used as a barostat for simulations of gases. We then turned our attention to a simple Lennard-Jones (LJ) fluid where a subset of particles has been treated as solute, without altering the interatomic interactions between the two types of LJ particles.
For these simulations, a Lennard-Jones potential with $\epsilon=0.9960$~kJ~mol$^{-1}$ and $\sigma=0.3405$~nm is used, which reproduces the phase diagram of argon,\cite{Pas20061991} and was used as a guide to choose the pressure and temperature conditions of our simulations.
Finally, we applied the osmotic pressure regulator to simulate the crystallisation of ice in the presence of an electrolyte solution using the mW water potential.\cite{molinero2009JPCB,demille2009JCP}

All MD simulations were conducted utilising OpenMM\cite{openmm} on a single GPU with mixed precision. The temperature was maintained through a Langevin integrator employing LFMiddle discretisation\cite{LangMiddle}. Given that the accurate determination of the crystallisation kinetics of ice is beyond the scope of this work, we utilised a small friction coefficient to reduce the effect of the thermostat on the crystal growth rate. 
We also performed supplementary verification for selected cases using the stochastic velocity rescaling thermostat\cite{csvr} to ensure that the results shown here are independent of the thermostat.

\subsubsection{Gas phase and ideal solutions}
Gas phase simulations containing 5,499 particles of argon in a $90 \times 90 \times 90$~nm cubic box were simulated at 300~K, which is well above the critical temperature of Ar (150~K), and its properties can be approximated with those of an ideal gas. The simulations were run using a timestep of 2~fs with a total simulation time of 1~$\mu$s. Three separate systems were run using a slab, cylinder, or sphere geometry for the osmotic pressure regulator, where every argon atom counted as a `solute' particle on which the membrane would act. The membrane fluctuated with a spring constant $K$ of 100~kJ~mol$^{-1}$~nm$^{-2}$; a compute interval of 1000~steps; a sample length for the rolling average of 10, and a relaxation time $\tau$ of 1.0~ps.

The simulations of ideal solutions were performed with the same Lennard-Jones potential but at a temperature of 100~K, which is well above the melting temperature (82~K), so that the application of any pressure would not result in the crystallisation of the system. Three separate systems were created for the slab, cylinder and sphere geometries: the slab geometry had a box size of $7 \times 7 \times 20.6$~nm, with the membrane radius defined along the $z$ axis; the cylinder geometry had a box size of $15 \times 15 \times 4.5$~nm, with the membrane radius defined in the $xy$ plane such that the cylinder runs parallel to the $z$ axis; the sphere geometry has a box size of $10 \times 10 \times 10$~nm. Each system contained 20,000 argon particles and was equilibrated at 100~K in the anisotropic NPT ensemble using the Monte Carlo barostat within OpenMM.\cite{openmm} 
Various amounts of argon were relabeled to allow the osmotic membrane to act only on a subset of the particles. In this context, argon is referred to as the solvent, and the relabeled argon is the solute. The number of solute atoms in the simulation ranged between 50 and 2,000.
The solute-solvent, solvent-solvent and solute-solute interactions were described using the same LJ parameters, effectively creating an ideal solution.

\subsubsection{Crystal growth dynamics}

The coarse-grained mW water model\cite{molinero2009JPCB} is used to model the crystal growth dynamics of ice's \hkl{001} basal surface. The force-field parameters for the interactions of sodium and chloride ions with the mW water model are taken from Ref.~\onlinecite{demille2009JCP}.
The forcefield ensures that the pairing distribution functions between the two ions and the mW water reproduce those obtained from simulations using the TIP/4P-EW model, and it has a reported melting temperature around 276~K.\cite{molinero2009JPCB,demille2009JCP} 
Experimentally, ice growth rates are relatively slow on MD time scales because forming ice layers involves a reorganization of the hydrogen bonding network. 
The mW model provides faster and somewhat unrealistic ice growth rates, as it is a coarse-grained model that treats each water molecule as a single particle, for which no hydrogen bond reorientation is required\cite{shi_salvalaglio}. 
We exploit this feature to our advantage, as faster growth rates allow us to accurately quantify the effect of including the constant osmotic pressure regulator developed in this work. 
However, to compute growth rates representative of real ice, a more accurate water model is needed, such as SPC/Fw\cite{wu2006JCP}, TIP4P\cite{abascal2005JCP}, \emph{etc}.

\begin{figure}[htbp]
    \centering
    \includegraphics[width=1.0\linewidth]{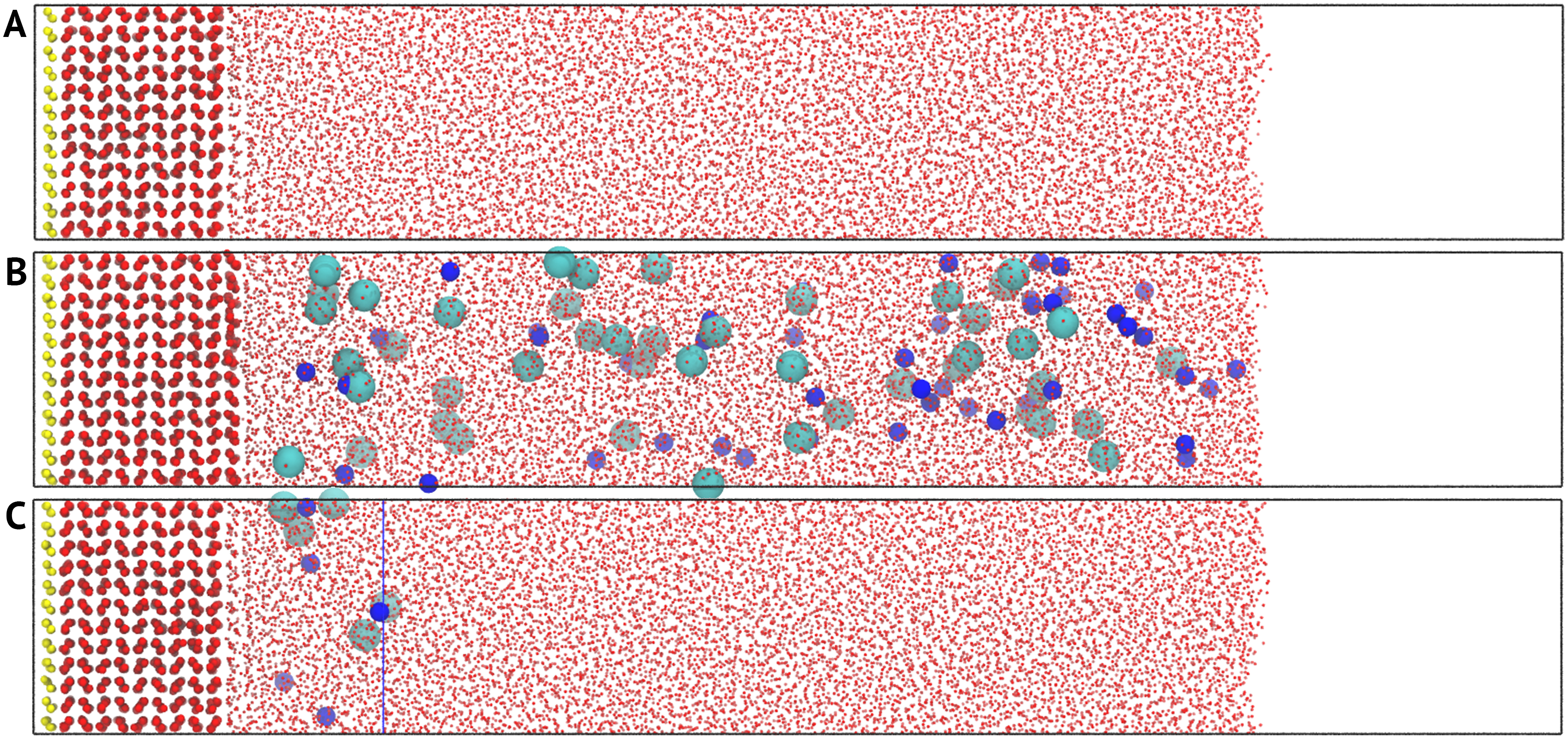}
    \caption{Graphical representation of the initial configurations of the three types of systems used to investigate the growth rate of the \hkl{001} basal plane of ice. Each system is aligned such that the surface normal is perpendicular to the \emph{z} axis. The crystalline water particles are depicted in red with a scaled van der Waals radius of 0.3. The non-crystalline water particles are depicted as red points. The bottom two layers of the crystalline water particles depicted in yellow represent the layer of restrained water particles. The sodium and chlorine ions are depicted in dark blue and cyan, respectively, with scaled van der Waals radii of 1.4. {\bf{A}} Pure water system, with no sodium or chlorine solute. {\bf{B}} System initialised with 46 sodium and 46 chlorine ions in the liquid phase, resulting in an osmotic pressure of 10~bar. {\bf{C}} System initialised with 5 sodium and 5 chlorine ions in the liquid phase above the surface. The blue vertical line indicates the initial location of the planar osmotic regulator, resulting in an osmotic pressure of 10~bar.}
    \label{fig:growth_renders}
\end{figure}

An orthorhombic supercell of the \hkl{001} basal surface of ice was constructed such that the surface normal was perpendicular to the \emph{z} axis of the cell. A fully flexible Monte Carlo barostat\cite{MCB1,MCB2} was used to equilibrate the cell at 273~K to determine the equilibrated \emph{x} and \emph{y} lattice parameters. An orthorhombic cell was constructed as 46.002 and 39.840~\AA{}~in the \emph{x} and \emph{y} directions, respectively, with 24 layers of water (2592 particles) frozen initially. 
The surface was then hydrated with 12050 particles of water at a density of $\approx$1 g/mL. The \emph{z} component of the cell was set to 300~\AA{}~to generate a section of vacuum between the liquid and the bottom of the periodic copy of the frozen section. The resulting cell was allowed to equilibrate for 0.5~ns in the NVT ensemble, and used as the starting point for the crystal growth simulations depicted in 
Figure~\ref{fig:growth_renders}. Simulations of the crystal growth were performed in the NVT ensemble at 273~K. The bottom two layers of the ice were restrained to the lattice positions using harmonic potentials to prevent drift of the frozen slab. The vacuum above the liquid is used to limit the growth to one direction and to accommodate the change in density whilst growing. 

The local ordering of water particles in the molecular dynamics simulations was quantified using a local version of the Steinhardt order parameters\cite{steinhardt1983PRB}, specifically the $q_6$ bond-orientational order parameter. This method characterises the local structure and symmetry around each particle by analysing the spherical harmonics of the bond orientations between neighbouring particles.\cite{moroni_interplay_2005,li_nucleation_2009,li_homogeneous_2011} The local $q_6$ parameter was calculated for each particle using a cutoff radius of 3.2~\AA{} to define the first coordination shell. The bond-orientational order parameter is defined as:
\begin{equation}
q_6(i) = \sqrt{\frac{4\pi}{13} \sum_{m=-6}^{6} |q_{6m}(i)|^2}
\end{equation}
\noindent where $q_{6m}(i)$ represents the complex spherical harmonics of order 6 and component $m$, averaged over all neighbors $j$ of atom $i$:
\begin{equation}
q_{6m}(i) = \frac{1}{N(i)} \sum_{j \in N(i)} Y_{6m}(\hat{\mathbf{r}}_{ij})
\end{equation}
To identify ordered particles, the dot product between the $q_6$ vectors of neighbouring particles was calculated:
\begin{equation}
\hat{q}_6 \cdot \hat{q}_6' = \frac{\sum_{m=-6}^{6} q_{6m}(i) \cdot q_{6m}^*(j)}{|q_6(i)| \cdot |q_6(j)|}.
\end{equation}
\noindent This normalised dot product serves as a measure of the correlation between local environments, with values above 0.5 indicating significant structural ordering. The degree of ordering in the system was quantified by summing the water particles with a correlation above 0.5.

\section{Results and Discussion}

The osmotic regulator is first validated \emph{via} simple systems that can be modelled analytically. In this way, the fundamental aspects of the regulator can be verified to reproduce predicted equilibrium solute densities/molar fractions, alongside investigating the effects of $\tau$, CI, SL, and $\kappa$ on the fluctuations and performance of the membrane. Secondly, the effects of the membrane on modelling non-equilibrium dynamics are investigated for the growth of ice in electrolyte solutions, demonstrating its practical utility for maintaining constant chemical potential conditions during crystallisation processes.

\subsection{Model systems}
\begin{figure}[htbp]
    \centering
    \includegraphics[width=0.8\columnwidth]{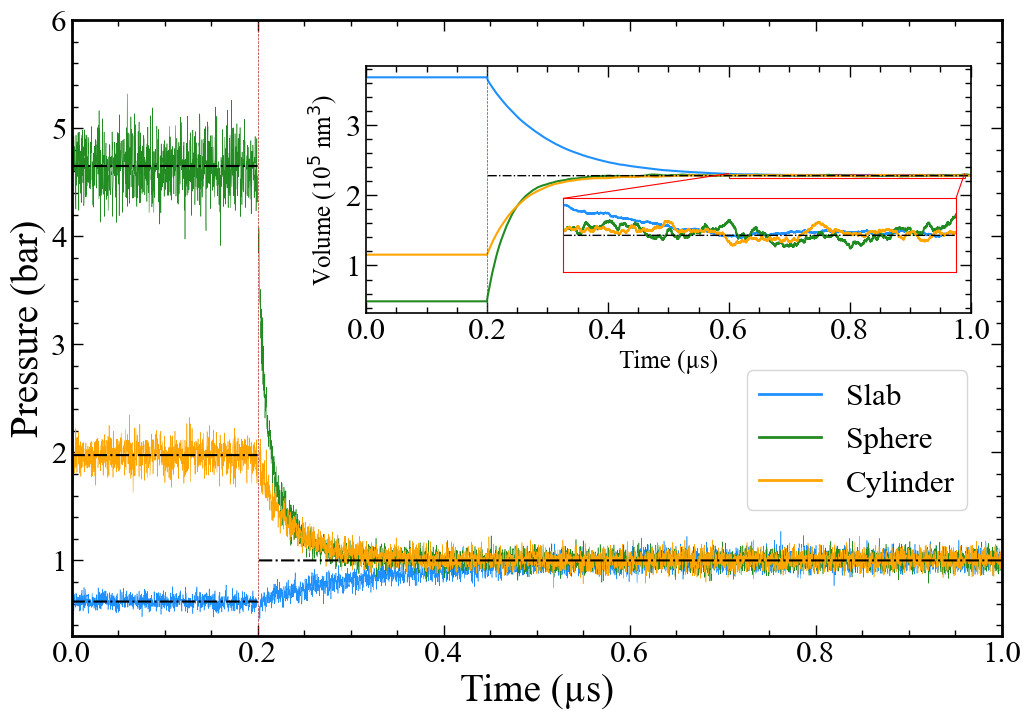}
    \caption{Comparison of the osmotic membrane's pressure and volume (inset) fluctuations. 
    For the first 200~ns, the membrane did not exert any force on the solute, only recording the pressure and volume. After that, the osmotic membrane was switched on to maintain an external osmotic pressure of 1~bar, with each particle counting as a `solute' particle for the membrane to act on, without any solvent. The horizontal black dot-dash lines represent the input target pressure of 1~bar and the corresponding volume that 5499 particles should occupy at 1~bar according to the ideal gas law.}
    \label{fig:compare_vacuum_big}
\end{figure}

\begin{figure}[htbp]
    \centering
    \includegraphics[width=0.5\columnwidth]{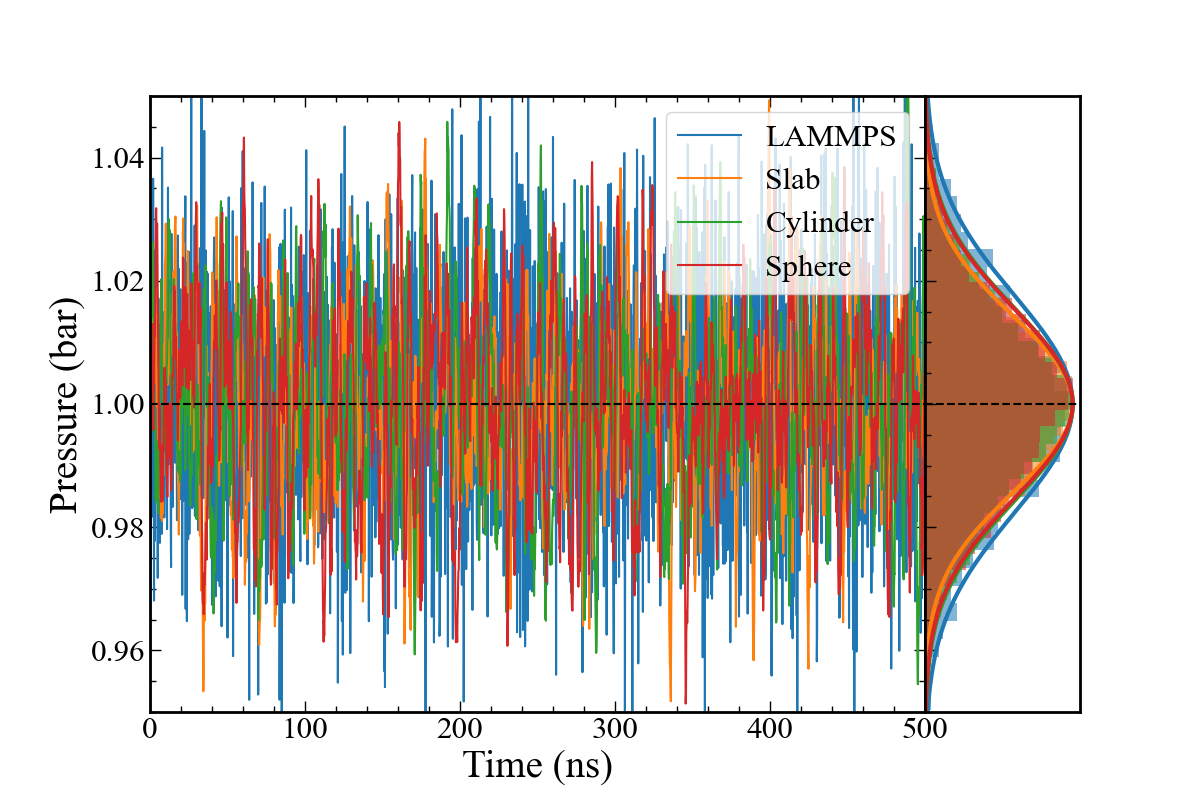}%
    \hspace{-0.5cm}
    \includegraphics[width=0.5\columnwidth]{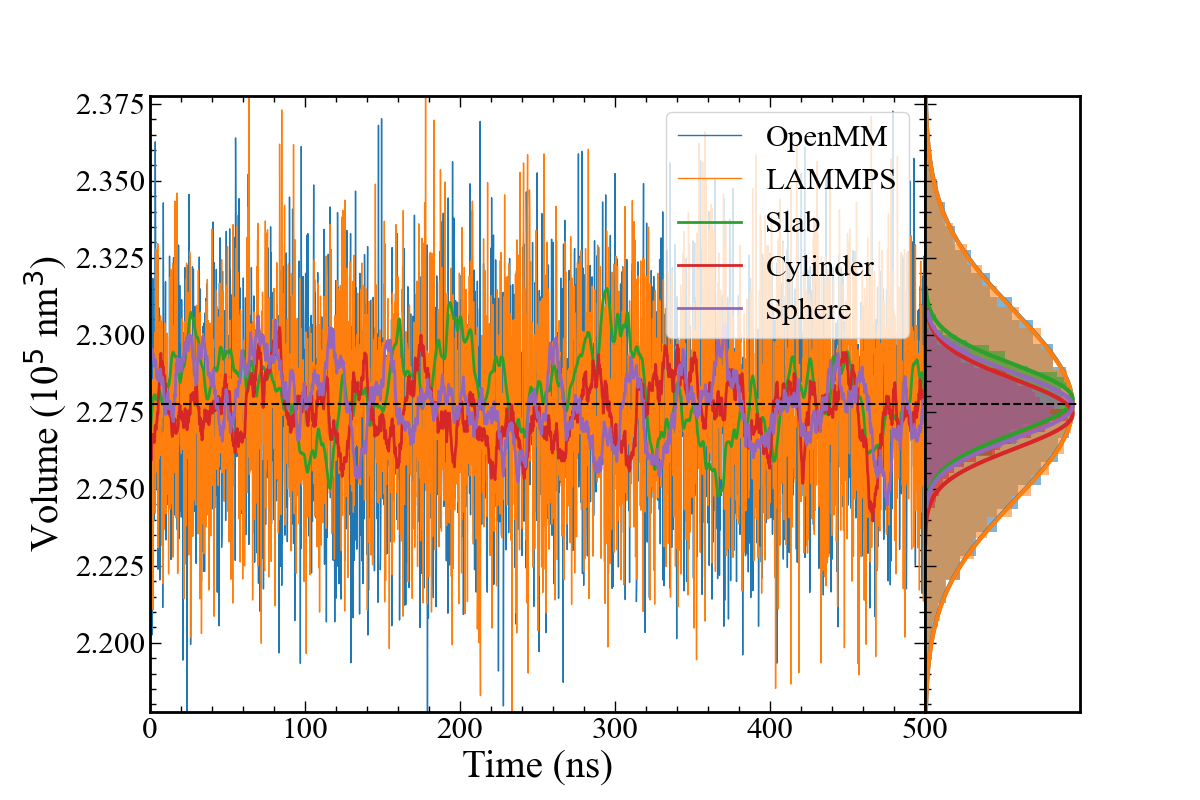}%
    \caption{Fluctuations in the pressure and volume output by the osmotic membrane for the ideal gas system described in the methods section over the course of 500~ns. The only difference is that an isothermal compressibility of 0.1~bar$^{-1}$ was used to match LAMMPS. Presented alongside are the pressure and gas fluctuations computed for the same system using the Berendsen barostat in LAMMPS (with an isothermal compressibility of 0.1~bar$^{-1}$) and the Monte Carlo barostat in OpenMM. The barostat within OpenMM does not compute the pressure and so can't be included when comparing the pressure fluctuations.}
    \label{fig:press_vol_fluctuations}
\end{figure}

For clarity, the starting radius of the osmotic membrane was varied between the slab, cylinder, and sphere geometries such that the initial pressures (and volumes) were distinct from one another, as demonstrated over the first 200~ns region in Figure \ref{fig:compare_vacuum_big} where the membrane was not being rescaled. Upon turning on the osmotic pressure regulator at 200~ns such that $R$ is scaled via Equation \ref{eq:berendsen}, the pressure and volume of each container's geometry converge to the input pressure of 1~bar and the corresponding volume given by $nRT/P$ from the ideal gas law. The fluctuations in the pressure and volume computed by the osmotic membrane are compared with those computed using the Monte-Carlo barostat in OpenMM\cite{openmm} and the Berendsen barostat as implemented in LAMMPS\cite{thompson2022CPC} in Figure \ref{fig:press_vol_fluctuations}.  
\noindent As expected, the fluctuations produced by the osmotic membrane do not reproduce those seen by the barostats, even when the isothermal compressibility $\kappa_T$ is identical.
Instead, the magnitude of the fluctuations is controlled by the assortment of tunable parameters that are included in the osmotic pressure regulator algorithm (CI, SL, $\kappa_T$, and $\tau$). 
Importantly, Figure~\ref{fig:press_vol_fluctuations} shows that the fitted Gaussian distributions of the pressure and volume are centred around the expected values. This validates the idea that the pressure can be computed as the sum of the restraining forces that the membrane applies to the particles divided by the membrane's surface area.\cite{luo_roux,yoshida2017JCP}
Even more crucial is that the distribution of the volumes for the various geometries is centred around the same value when computed for a cubic box \emph{via} the Monte-Carlo and Berendsen barostats, thus validating the integrals for the flat-bottom harmonic restraint's volume in 1-,2- and 3-dimensions (see appendix for details).

\begin{figure}[htb]
    \centering
    \includegraphics[width=0.75\columnwidth]{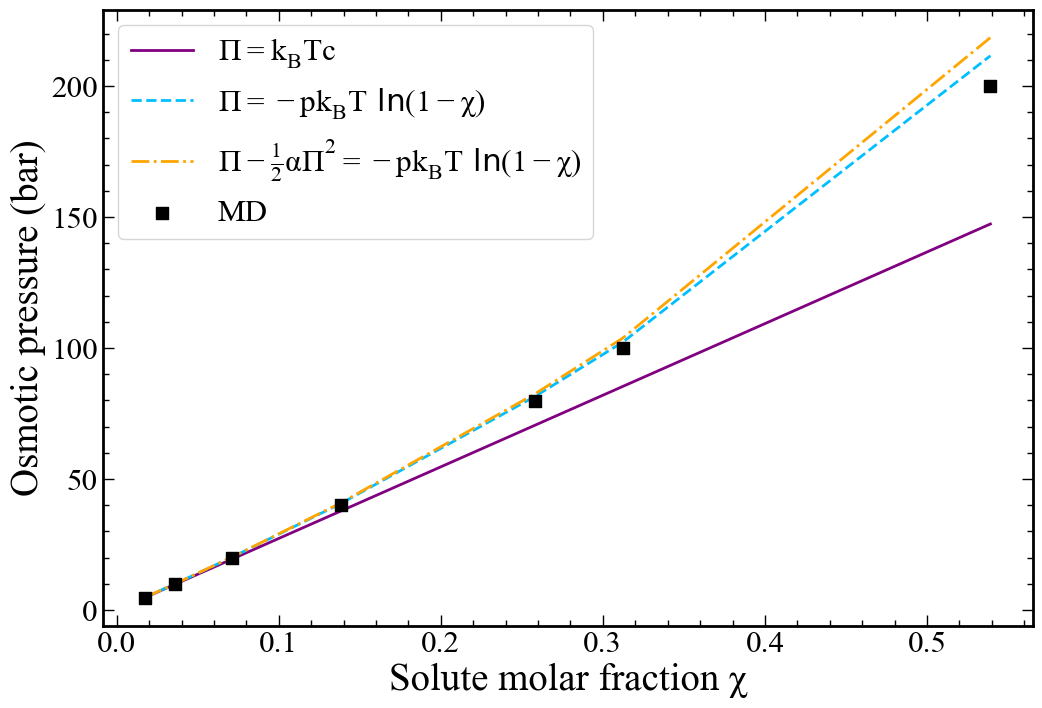}
    \caption{Comparison of various levels of theory for analytically predicting the osmotic pressure with the values calculated \emph{via} MD for liquid Argon. The MD data was generated from simulations of various amounts of solute inside a solvent, with each simulation having the same number of particles in total (20000). Each simulation was run for 200~ns at 100~K using a timestep of 2~fs in a box $70 \times 70 \times \approx 206$~\AA. The osmotic pressure was maintained using a slab-like membrane with a spring constant of 100 $\mathrm{kJ}~\mathrm{mol}^{-1}~\mathrm{nm}^{-2}$; a sample length of 100 previous computes; a relaxation time ($\tau$) of 1.0~ps; and a compute-interval of 100~steps. The averaged value of the converged osmotic pressure was then plotted against the molar fraction of the solute, which was computed using the average converged volume of the membrane. The three different levels of theory are shown in the legend and explained in Equations \ref{eq:osmotic_1}, \ref{eq:osmotic_2}, and \ref{eq:osmotic_3}.}
    \label{fig:osm_v_conc}
\end{figure}

The next point of verification is to ensure that the osmotic membrane maintains an expected solute molar fraction as calculated by the standard physical chemistry equations for an ideal solution (Eq \ref{eq:osmotic_1}, \ref{eq:osmotic_2}, and \ref{eq:osmotic_3}). To this aim, we used the same forcefield for argon but reduced the temperature to 100~K, where it is liquid, and tagged various amounts of atoms as solute, while keeping the interatomic interactions unchanged. The simulations were set up using the slab geometry with two parallel restraining surfaces for the osmotic membrane. The target osmotic pressure was set at 7 different pressures: 5, 10, 20, 40, 100, and 200~bar, and then the position of the osmotic membrane was allowed to equilibrate. The solute molar fraction was then computed using the averaged volume from the simulation. The osmotic pressure calculated \emph{via} MD was then compared against the three levels of theory for predicting the osmotic pressure of an ideal solution (Eqs. \ref{eq:osmotic_1}, \ref{eq:osmotic_2}, \ref{eq:osmotic_3}). Figure \ref{fig:osm_v_conc} depicts excellent agreement of the output of the osmotic membrane with the analytical predictions of the variation in osmotic pressure with solute concentration. A slight deviation between MD and analytical predictions is only seen when the molar fraction of the solute is greater than half ($\chi > 0.5$), where there is no clear distinction between solute and solvent. 

\begin{figure}[htbp]
    \includegraphics[clip,width=0.5\columnwidth]{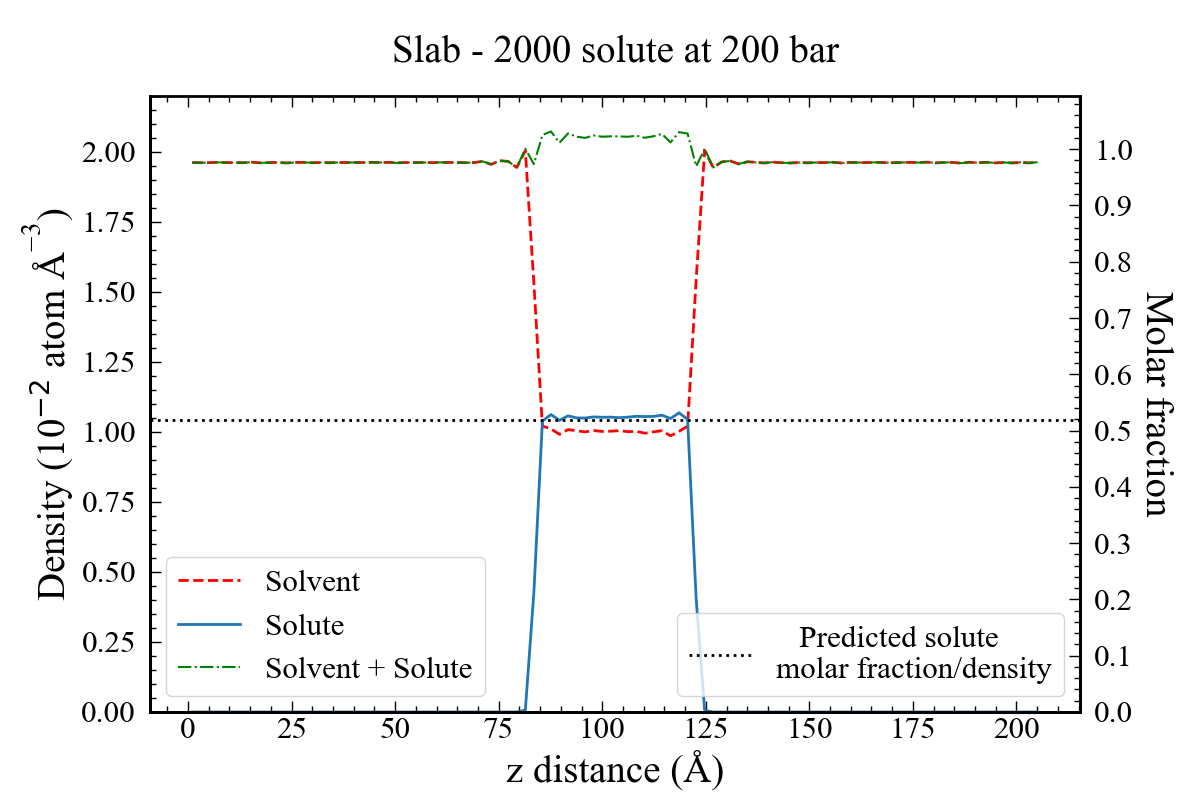}%
    \hspace{0.0cm}
    \includegraphics[clip,width=0.5\columnwidth]{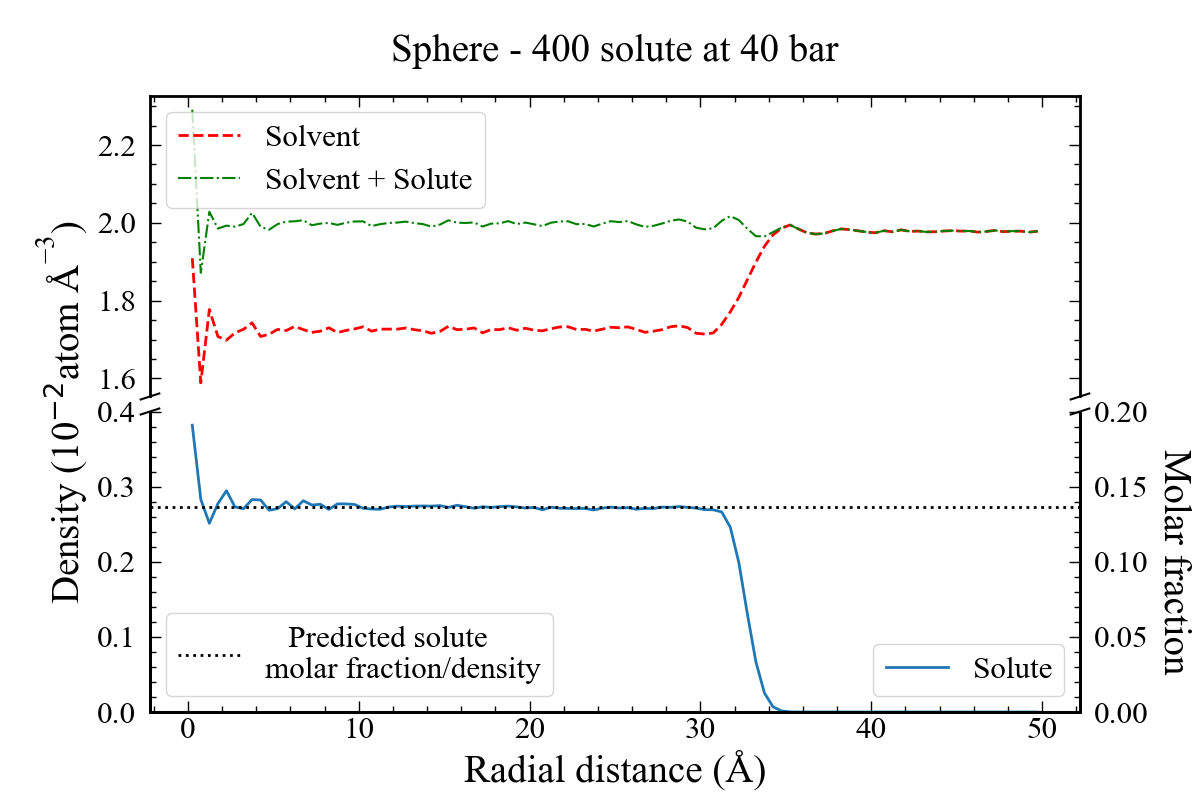}%
    \hspace{0.0cm}
    \includegraphics[clip,width=0.5\columnwidth]{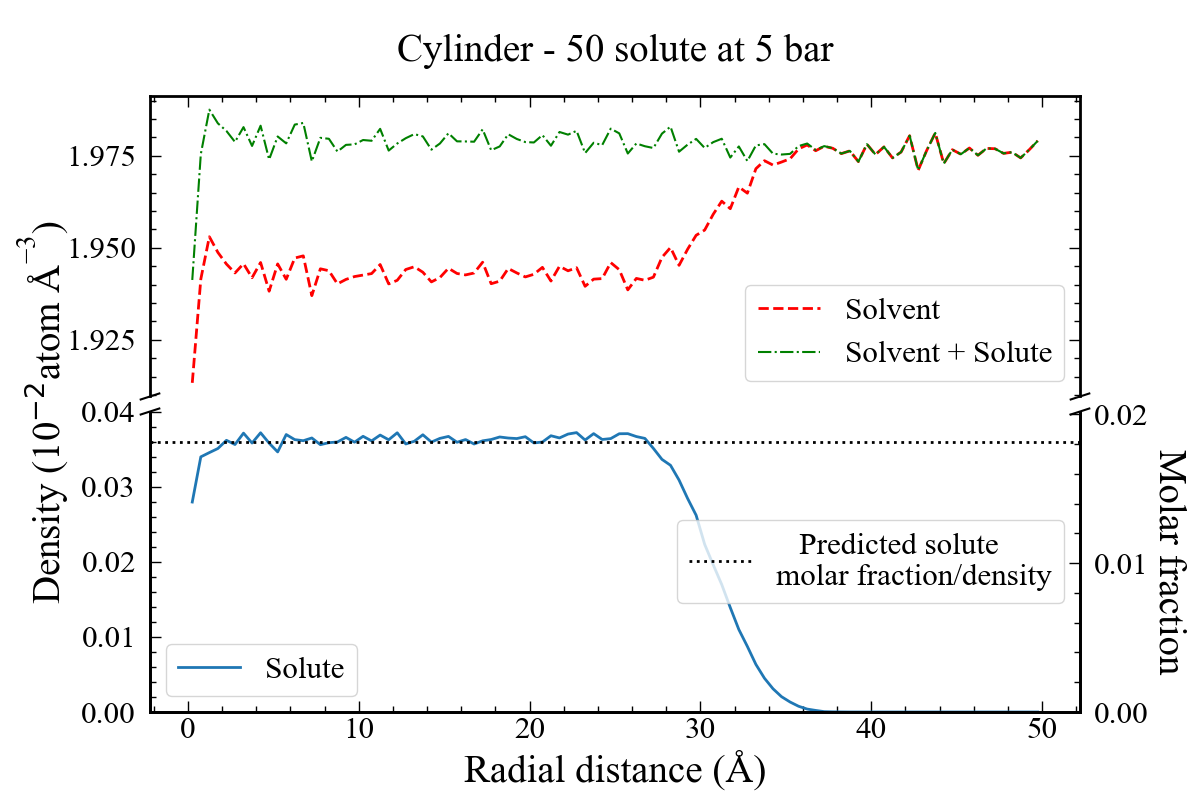}%
    \caption{One-Dimensional density profiles of the solute particles averaged over an MD trajectory while using either a slab, sphere, or cylindrical osmotic membrane. Each simulation was run for 200~ns at 100~K using a timestep of 2~fs with a total of 20000 particles (solute + solvent). The number of solute particles, the target osmotic pressure and the membrane geometry are presented in the title for each graph. The osmotic pressure was maintained with a spring constant of 10,000 $\mathrm{kJ}~\mathrm{mol}^{-1}~\mathrm{nm}^{-2}$; a sample length of the previous 1000 computes; a relaxation time $\tau$ of 1.0~ps; and a compute-interval of 1000~steps. The analytically predicted solute molar fraction is presented as a dashed horizontal line using the `Molar fraction' axis on the right, which was computed using Eq.~\ref{eq:osmotic_2}.}
    \label{fig:compare_analytic}
\end{figure}

Three further ideal solution systems were constructed to verify that the equilibrium volume produced by each membrane geometry had a corresponding solute density equal to the predicted value.
In order to highlight the flexibility of the methodology proposed here, we used 2000 solute particles at 200~bar of osmotic pressure in the slab geometry, 400 solute particles at 40~bar of osmotic pressure in the sphere geometry, and 50 solute particles at 5~bar of osmotic pressure in the cylinder geometry. 
Note that this cannot be easily done for the plane geometry, where only one restraining surface is used, and the volume occupied by the solute would be ill-defined.
The density of the solute is then averaged over 200~ns. In the case of the slab geometry, the density is averaged in the plane of the membrane along the orthogonal dimension ($z$ in this case). 
For the sphere and cylinder geometries, we computed the average radial densities from the axis of the cylinder or the centre of the sphere.
The densities of the solute, solvent, and solute $+$ solvent for each geometry are reported in Figure~\ref{fig:compare_analytic} and compared with the analytically predicted density from Equation~\ref{eq:osmotic_2}, which shows definitive agreement between the two. 

\begin{figure}[htbp]
    \centering
    \includegraphics[width=0.5\columnwidth]{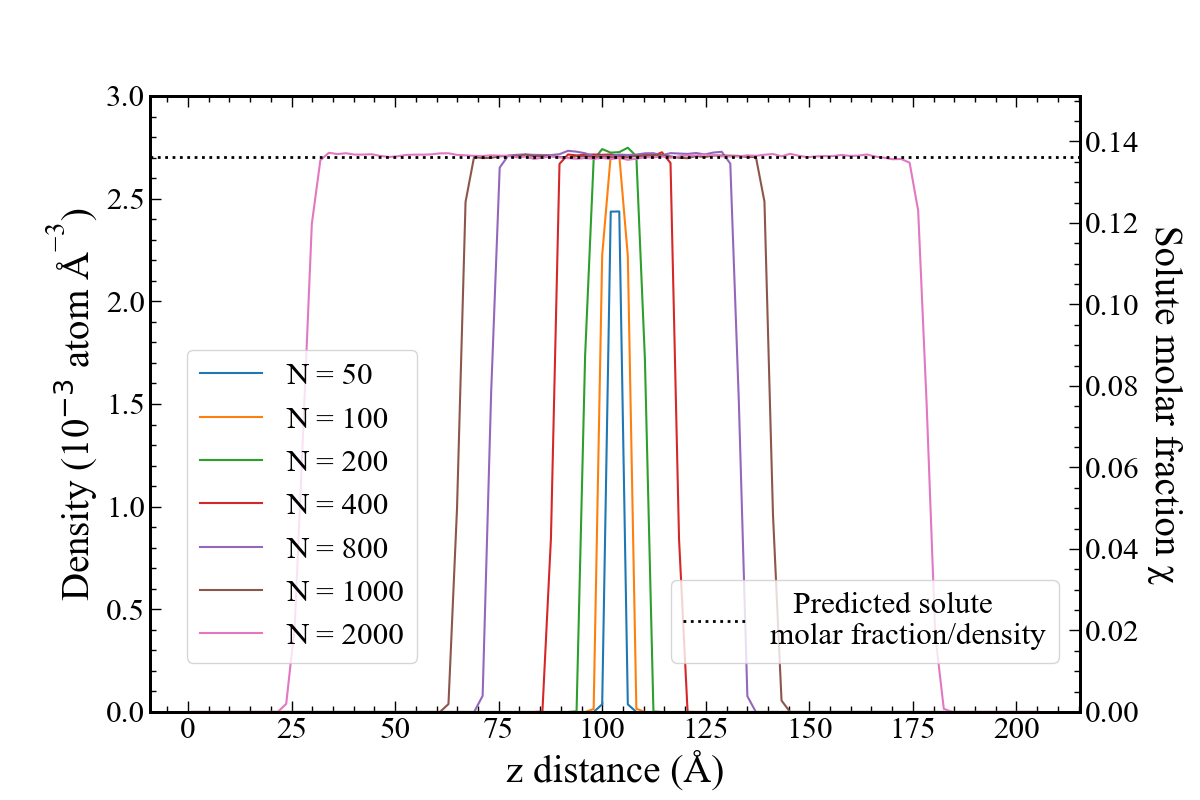}
    \hspace{-1cm}
    \includegraphics[width=0.5\columnwidth]{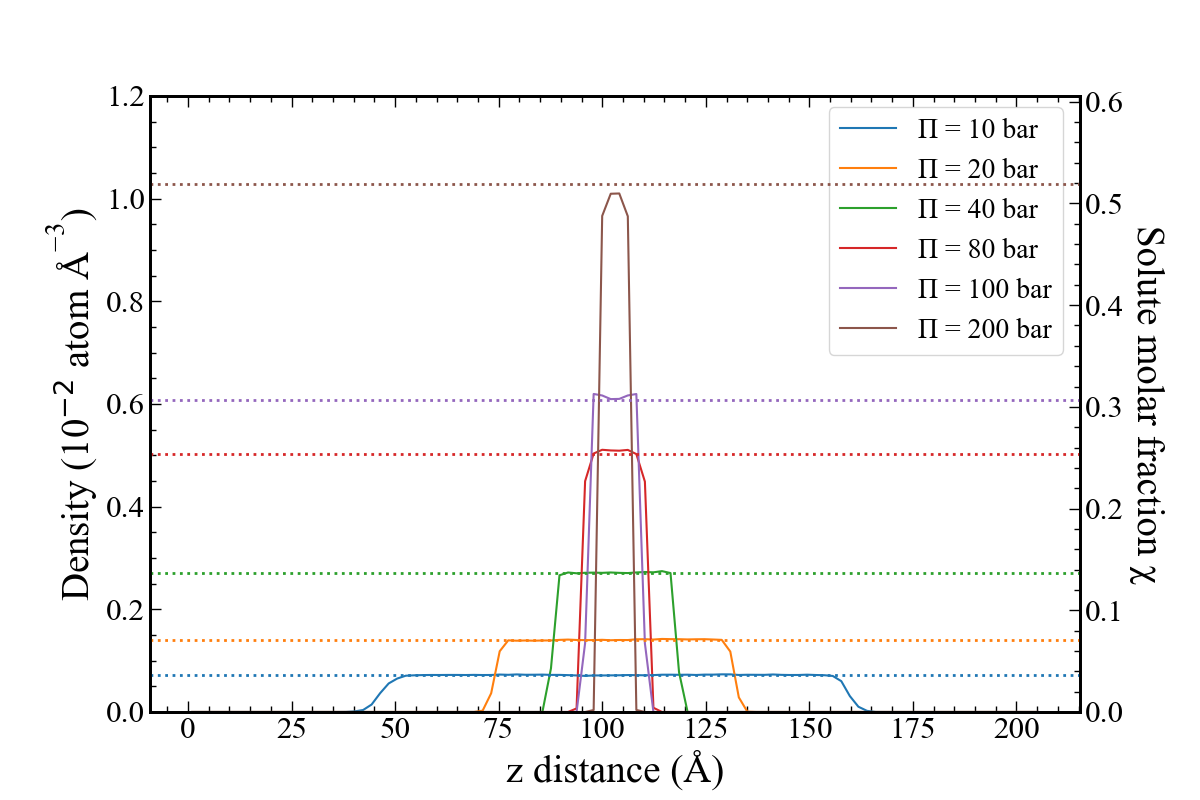}
    \caption{Comparison of the one-dimensional density profiles of various amounts of solute particles (N) in 20000 total particles (solute + solvent) along the \emph{z}-axis (top) and comparison of the one-dimensional density profiles of 400 solute particles in 20000 total particles (solute + solvent) along the \emph{z}-axis (bottom). The osmotic pressure was maintained by a fluctuating slab-like membrane in the \emph{xy} plane at 40~bar. The density profiles were averaged over 200~ns of NVT simulation at 100 K after 400~ns of equilibration with a timestep of 2~fs. The membrane fluctuated with a spring constant of 10,000 $\mathrm{kJ}~\mathrm{mol}^{-1}~\mathrm{nm}^{-2}$; a sample length of the 1000 previous computes; a relaxation time $\tau$ of 1.0~ps; and a compute-interval of 1,000~steps. The analytically predicted solute molar fraction is presented as a dashed black horizontal line using the `Solute molar fraction' axis on the right, which was computed using Eq.~\ref{eq:osmotic_2}.}
    \label{fig:compare_solute_osmotic}
\end{figure}

Moreover, in Figure \ref{fig:compare_solute_osmotic} we demonstrate how the osmotic pressure regulator can produce the correct solute density for different amounts of solute particles at a chosen osmotic pressure, 40~bar in this case. 
Notably, while this works well for large amounts of solute, and the membrane is able to maintain a uniform density throughout the entire volume, this is not the case for 50 solute particles.
This is due to the fact that the area of the membrane in our simulation box was too large and there were not enough solute atoms to produce the desired osmotic pressure even at $R=0$.

\begin{figure}[htbp]
    \centering
    \begin{minipage}{0.48\textwidth}
        \centering
        \text{$\tau$}
        \includegraphics[width=\textwidth]{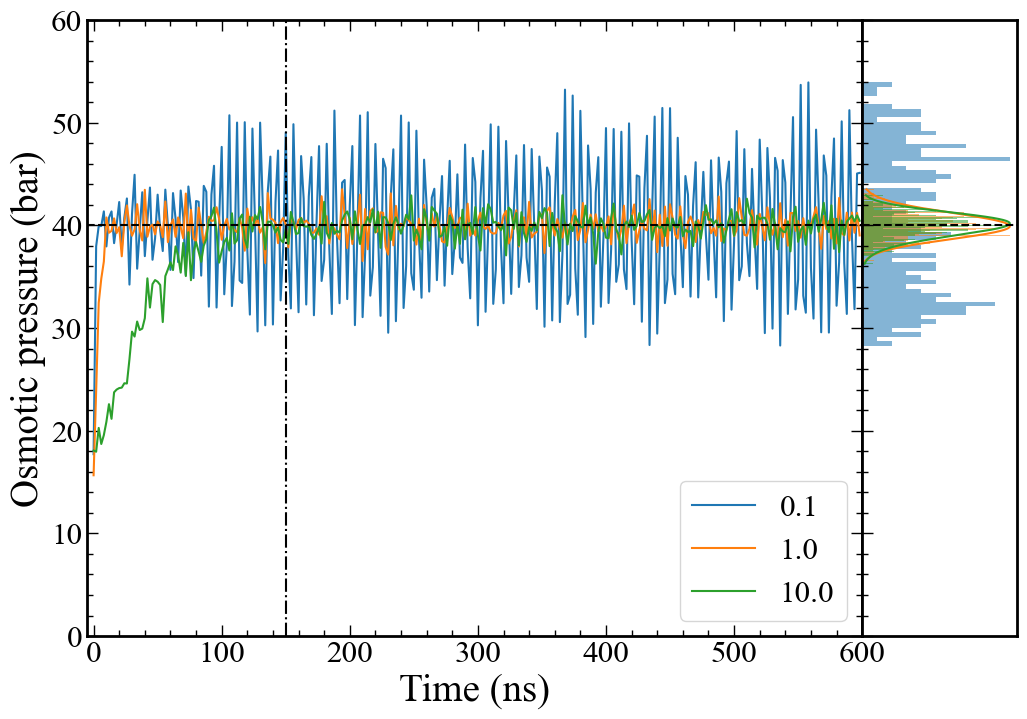}
        \label{fig:sub1}
    \end{minipage}
    \begin{minipage}{0.48\textwidth}
        \centering
        \text{Compute interval (CI)}
        \includegraphics[width=\textwidth]{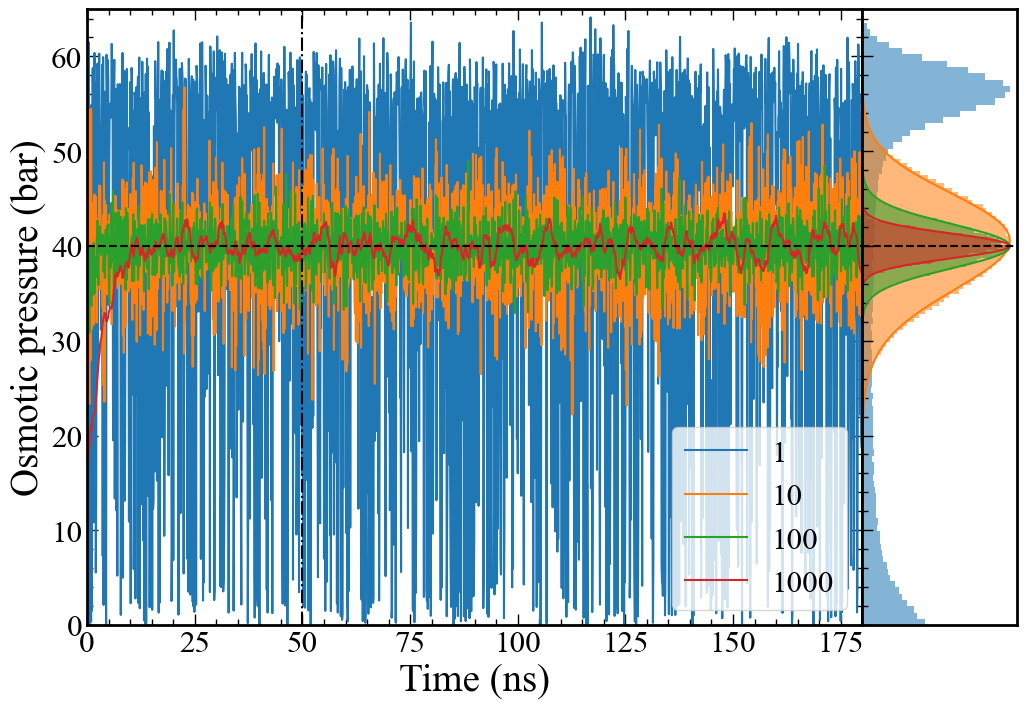}
        \label{fig:sub2}
    \end{minipage}
    \vspace{-0.5cm}
    \begin{minipage}{0.48\textwidth}
        \centering
        \text{Sample length (SL)}
        \includegraphics[width=\textwidth]{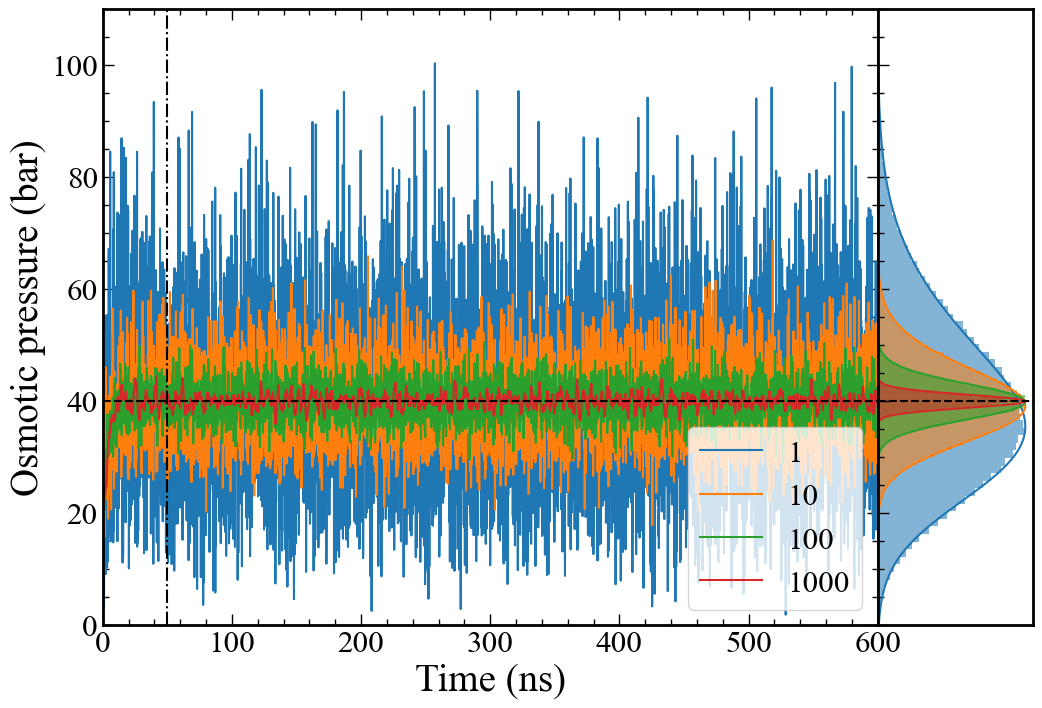}
        \label{fig:sub3}
    \end{minipage}
    \begin{minipage}{0.48\textwidth}
        \centering
        \text{$\kappa$}
        \includegraphics[width=\textwidth]{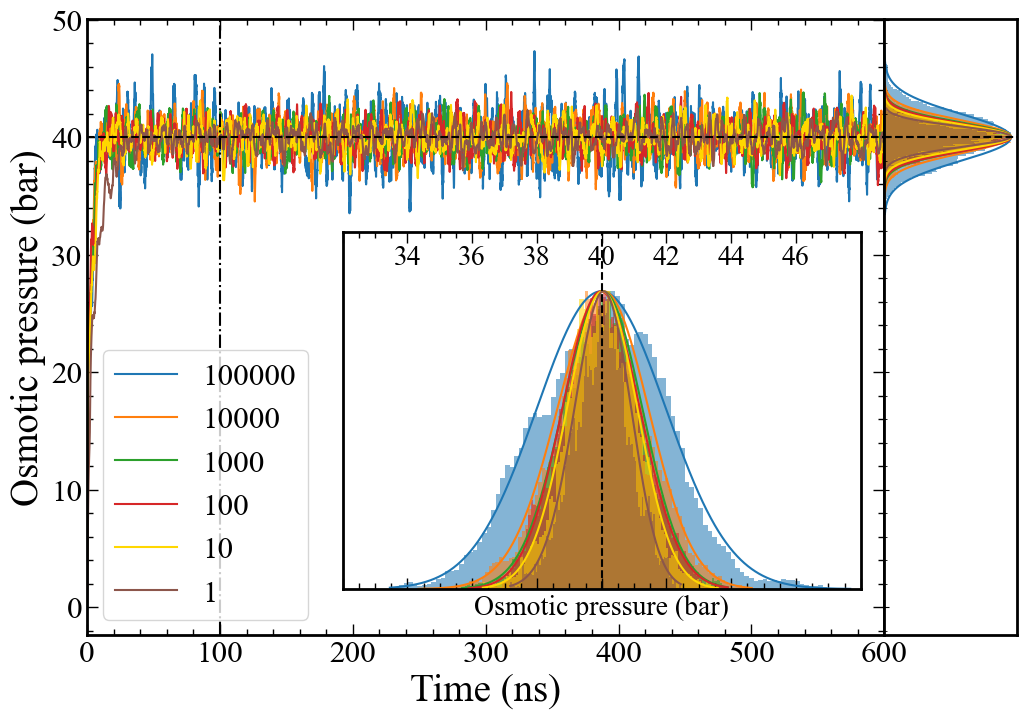}
        \label{fig:sub4}
    \end{minipage}
    \caption{Comparison of the osmotic pressure fluctuations ($\Pi(\mathrm{GCMD})$) for a system containing 400 solute particles in 20000 total particles (solute + solvent) in the NVT ensemble at 100~K with a timestep of 2~fs. The osmotic pressure was maintained by a fluctuating slab-like membrane in the \emph{xy} plane at 40~bar. The title of each graph denotes the regulator parameter changed according to the legend. Unless changed, the values for the osmotic regulator parameters were set to 1~ps for $\tau$, $1000$ computes for SL, 1000~steps for CI, and $1000$~$\mathrm{kJ}~\mathrm{mol}^{-1}~\mathrm{nm}^{-2}$ for $\kappa_T$. The horizontal black dashed line represents the input target osmotic pressure of 40~bar. After equilibration (denoted by the vertical dot-dash line), the osmotic pressure fluctuations are shown on the right as a fitted (where appropriate) Gaussian distribution normalised by height using the same representative colours.}
    \label{fig:param_grid}
\end{figure}

The fluctuations and convergence of the osmotic pressure depend on several parameters that control the regulator. For instance, the magnitude of $\tau$ is directly proportional to the magnitude of the fluctuations about the target osmotic pressure. The fluctuations are investigated for $\tau=$~0.1, 1.0, and 10.0~ps, the results of which are presented in Figure~\ref{fig:param_grid}. A value of 0.1~ps produces excessively large fluctuations with an average centred about the external osmotic pressure of 40~bar, but the distribution is not Gaussian with a higher density of points at $\approx$32 and $\approx$48 bar. Values 1.0 and 10.0~ps produce more reasonable Gaussian distributions about 40~bar, with the value of 1.0~ps converging to 40~bar faster. The effects of the CI, SL and $\kappa$ parameters on the osmotic pressure fluctuations are shown in Figure ~\ref{fig:param_grid}. For the CI parameter, a value of 1 causes rapid fluctuations in the osmotic pressure as the membrane bounces back and forth wildly. Values 10, 100 and 1000 produce non-skewed Gaussians centred about the input target osmotic pressure of 40~bar, with the distribution getting narrower as the CI value increases. Similar to the CI parameter, an SL value of 1 produces too large fluctuations, resulting in a skewed distribution. However, 10, 100 and 1000 SL values produce non-skewed Gaussians with the correct average of 40~bar, with the width of the distribution decreasing with increasing SL. Finally, the fluctuations in the osmotic pressure are relatively insensitive to the magnitude of $\kappa_T$.

\subsection{Ice growth}
\begin{figure}[htb]
    \centering
    \includegraphics[width=0.8\linewidth]{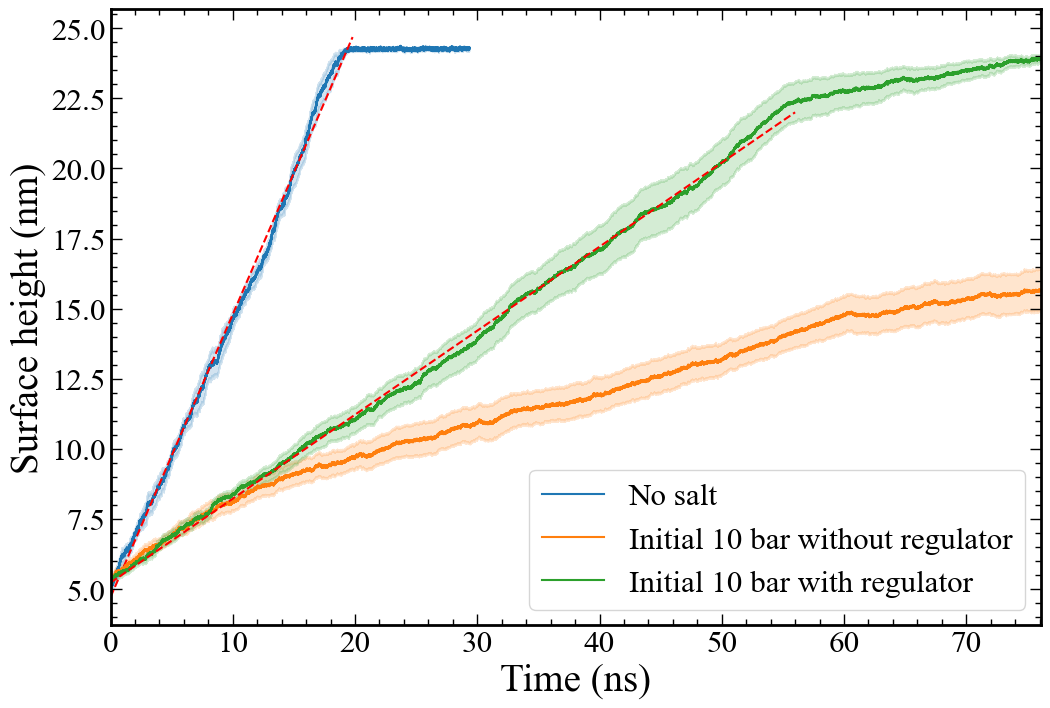}
    \caption{Comparison of the position of ice front of the \hkl{001} basal plane recorded during an NVT run at 2~K below the mW water melting temperature. For each system, the line represents the average over 15 replicates, with the shaded area denoting the 95\% confidence interval.
    The ice growth rate can be obtained from the slope of the linear part of the curves (red dashed lines).}
    \label{fig:compare_reg_growth}
\end{figure}

The growth of the \hkl{001} basal plane of ice was investigated for each of the three systems described in Figure~\ref{fig:growth_renders}, with results presented as a function of simulation time in Figure~\ref{fig:compare_reg_growth}. The data depicts the average over 15 replicates, with 95\% confidence intervals shown as shaded regions about the average. The effective height of the surface was quantified by dividing the total number of water particles with $\hat{q}_6 \cdot \hat{q}_6' > 0.5$ by the number of water particles that construct a layer in the bulk cell with the same \emph{x} and \emph{y} lattice parameters. This calculation produced the number of layers in the solid, which was then multiplied by the spacing between layers to obtain the surface height.

The system of pure water (Figure~\ref{fig:growth_renders}{\bf{A}}) and the solute system with the osmotic regulator (Figure~\ref{fig:growth_renders}{\bf{C}}) were simulated until ice growth plateaued, when virtually all available water had crystallised except for a small portion atop the crystal. The unregulated solute system (Figure~\ref{fig:growth_renders}{\bf{B}}) was simulated for an equivalent duration as the regulated system. It was observed that the pure water system demonstrated a linear growth rate, as did the solute system with the regulator. The greater variation in the replicates of the solute system was attributed to the incorporation of solute into the crystal at varying degrees between replicates. This effect was demonstrated even more prominently in the system without the regulator, where growth kinetics were complicated by competing processes. Theoretically, the growth rate would plateau as the concentration of solute in the liquid reached the point at which the melting temperature decreased below the simulation temperature. However, since the salt was readily incorporated into the crystal as it grew, the effective concentration of salt in solution decreased accordingly. As such, the observed non-linear growth rate arose from the complex interplay between increasing solute concentration from impurity-free growth and decreasing solute concentration as impurities were incorporated into the crystal\cite{sosso_7sins}.

The simulation temperature of 273~K was specifically chosen to limit the growth rate, being only slightly below the melting temperature, which with our implementation of the potential in openMM is approximately 277~K, in good agreement with the values of 275~K and 276~K reported in the literature\cite{molinero2009JPCB,demille2009JCP}. 
This approach reduced the amount of solute incorporated into the crystal, better demonstrating the increase in solute concentration in the liquid phase during crystal growth. The behaviour observed in the non-regulated system is entirely artificial and a consequence of the limited sizes available to MD simulations. In reality, as crystals grow, interfacial solute particles freely diffuse into the broader solvent reservoir, maintaining consistent concentration levels throughout the growth process. The implemented osmotic regulator provides an effective solution to this simulation artifact, enabling the determination of growth rates that more accurately reflect real-world conditions. 

\begin{figure}[htbp]
    \centering
    \includegraphics[width=1.0\columnwidth]{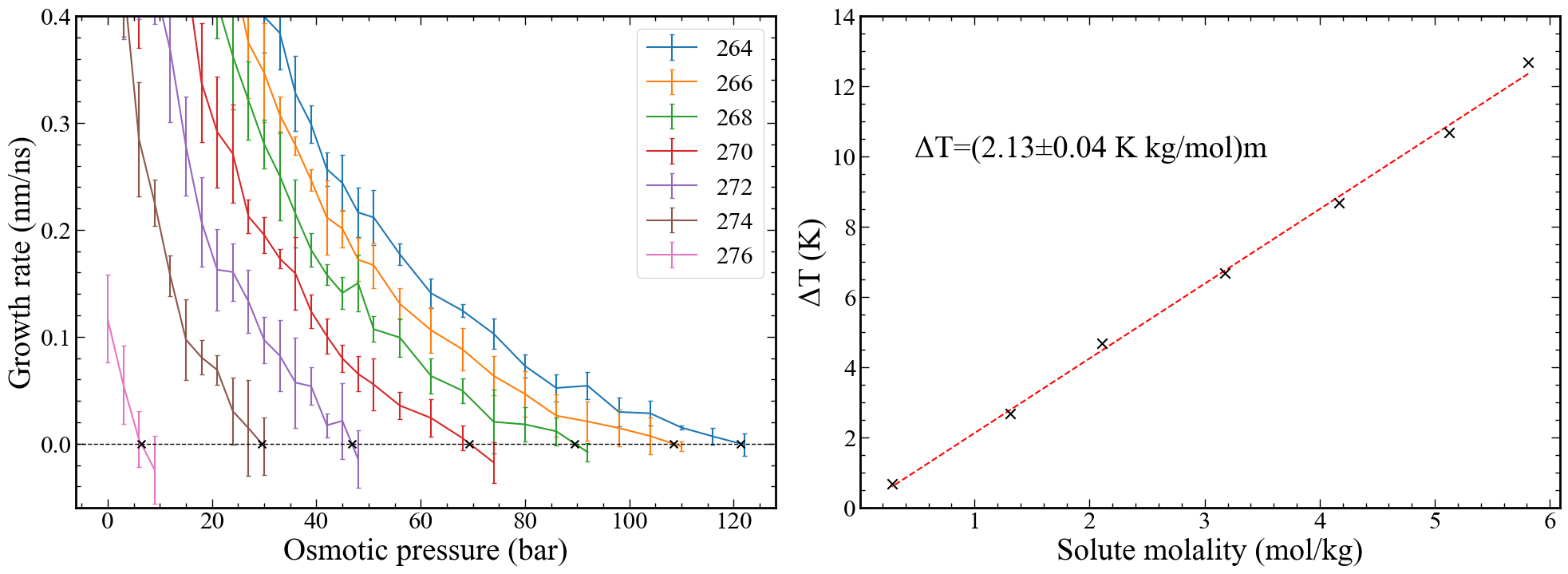}
    \caption{Left: Growth rate of the basal plane of ice as a function of solute osmotic pressure. The legend denotes the various temperatures at which the simulations were conducted, with the error bars denoting 95\% confidence intervals from five replicates. Right: The freezing point depression corresponding to the solute molality of the simulations. The equation resulting from the linear regression is presented in the body of the graph.}
    \label{fig:colligative}
\end{figure}

In order to further validate our methodology, we used the osmotic pressure regulator to investigate the freezing point depression of ice using the mW water potential. 
To this aim, we constructed a set of water ice interfaces similarly to those shown in Figure~\ref{fig:growth_renders}{\bf{C}} with solute concentrations equivalent to osmotic pressures in the range of $0-200$~bar. 
Each system was then simulated with the regulator set to the initial osmotic pressure and at temperatures between $264-276$~K in 2~K increments. Each configuration was simulated for five replicates, and the growth rate was computed at each temperature by fitting the line to the position of the ice interface as a function of the simulation time, as shown before.
The osmotic pressure required to change the ice melting temperature to the simulation temperature was then obtained using linear interpolation between the first point where the growth is negative and the previous point. 
The osmotic pressure that corresponds to the crossover point was then converted
to molality by
\begin{equation}
    m=\frac{55.51~\mathrm{mol}~\mathrm{kg}^{-1}\cdot \chi_{solute}}{1-\chi_{solute}}
\end{equation}
\noindent where $m$ is the solute molality, $55.51~\mathrm{mol}~\mathrm{kg}^{-1}$ is the molality of pure water and $\chi_{solute}$ is the solute molar fraction, as computed using Equation~\ref{eq:osmotic_2}, which was shown to be valid for small molar fractions.
The reduction of the ice melting temperature ($\Delta$T) shows perfect linear dependency on the molality of the solution (shown in Figure~\ref{fig:colligative}), in agreement with the classical thermodynamic theory of the colligative properties of solutions
\begin{equation}
\Delta \mathrm{T}=K_f\cdot m
\end{equation}
where $K_f$ is a solvent-dependent proportionality constant referred to as the cryoscopic constant, and $m$ is the molality of the solution.
By ensuring that the fitting line passed through the origin of the axis, we could also improve our calculation of the mW water melting temperature to be $276.7$~K.
The computed cryoscopic constant for mW water is $2.13\pm0.04$~K~kg~mol$^{-1}$, which is in good agreement with the reported value for water of $1.86$~K~kg~mol$^{-1}$.\cite{scatchard1960JCP}
This is not surprising since the mW water potential has been designed to closely reproduce the water enthalpy of fusion,\cite{molinero2009JPCB} on which the cryoscopic constant depends, and it further validates the methodology developed here to dynamically control the osmotic pressure during MD using a Berendsen-like algorithm.

\section{Conclusions}
In conclusion, we presented here a novel Berendsen-like osmotic pressure regulation algorithm that provides an accessible and computationally efficient alternative to traditional GCMD methods. This approach circumvents the challenges associated with the insertion/deletion of particles by maintaining a constant, externally applied osmotic pressure, which effectively keeps the solute chemical potential constant. This is achieved \emph{via} flat-bottom harmonic potentials applied in various dimensions to produce geometric configurations (planar, slab, cylindrical, and spherical) that restrain particles to an analytically known volume corresponding to the externally applied osmotic pressure. The osmotic regulator has been demonstrated to successfully reproduce ideal gas law behaviour across different geometries, with pressure and volume fluctuations centred around the expected values. For ideal solutions, the regulator demonstrates consistency with Lewis' osmotic pressure formulation across a wide range of solute concentrations. The practical utility of the osmotic regulator is demonstrated through crystal growth simulations of the \hkl{001} surface of ice, where different growth rates are observed between regulated and unregulated systems. Unregulated simulations exhibit artificial non-linear growth rates arising from changes in the concentration occurring during the freezing process. 
In contrast, the regulated simulations maintain consistent, linear growth rates that better represent real-world conditions where crystals grow from an effectively infinite reservoir, even when the solute is partially incorporated in the growing crystal. 

The key benefits of the implementation described here include:
\begin{itemize}
    \item elimination of the computational complexity associated with particle insertion/deletion algorithms
    \item straightforward implementation in existing MD software (OpenMM\cite{openmm}) without requiring fundamental changes to core simulation engines
    \item flexible geometric configurations that can accommodate a variety of simulation setups
    \item intuitive parameter control based on established Berendsen barostat principles.
\end{itemize}
The osmotic regulator provides an efficient and user-friendly way to simulate a range of phenomena such as freezing, nucleation and growth of crystals at a constant chemical potential, overcoming some of the limitations that MD suffers from due to finite size effects\cite{sosso_7sins}.

\begin{acknowledgments}
BIA thanks the Australian Research Council for funding through grant FL180100087. DD acknowledges support by the U.S. National Science Foundation under Grant No. 2053235.
This work was supported by resources provided by the Pawsey Supercomputing
    Research Centre with funding from the Australian Government and the
    Government of Western Australia, as well as by the National Computational Infrastructure through an NCMAS allocation.  
\end{acknowledgments}

\section*{DATA AVAILABILITY}
The data that support the findings of this study are available from the corresponding author
upon reasonable request.
Installation instructions and example runs for utilising the osmotic pressure regulator within OpenMM are available on GitHub and Zenodo.\cite{opo_blake}


\section*{Appendix}
The effective volume imposed by the regulator and the effective surface area of the restraint to be used for the calculation of the osmotic pressure depend on the chosen geometry and the magnitude of the spring constant of the harmonic restraint. 
In this section, we show how this can be derived.
The restraint always has the functional form of a flat-bottom potential
\begin{equation}
U(r) = 
\begin{dcases*}
    0 & if  $\lvert r\rvert \leq$ R \\
    \frac{1}{2}K(r-R)^2                   & \text{otherwise}
\end{dcases*}
\end{equation}
where $r$ is the 1-D, 2-D or 3-D distance from the centre of the restraint in, $R$ is a parameter which determines the size of the flat-bottom potential, and $K$ is the strength of the harmonic restraint.

\subsection{1-Dimensional}
In the case of the 1-D restraint, the surface area of the regulator is simply the area of the simulation cell normal to the direction of the restraint for a planar restraint, and twice that for a membrane restraint where the osmotic pressure acts on both the top and bottom surface of the restraint.

While the volume of the plane restraint is ill-defined, due to the difficulty of assigning a fixed position for the solid surface, it is possible to compute the effective volume of a soft slab restraint by determining the effective thickness of the slab, which depends on $K$ and the temperature of the system. In fact, the softer the restraint, the more the particles can spill out of the restraint, making the effective thickness larger. 
Assuming the slab restraint is normal to $x$,
\begin{equation}
r(x) = \sqrt{(x-x_0)^2}
\end{equation}
\noindent the corresponding partition function $Q$ for the 1-D flat-bottomed harmonic restraint can be written as:
\begin{equation}
Q=\int_{-\infty}^{+\infty}\exp\bigg(-\frac{U(r)}{k_BT}\bigg)\ \mathrm{d}x\ \\.
\end{equation}
Therefore, the effective length $L$ along the $x$ axis that a particle can occupy at a given temperature $T$ and flat-bottom region $R$ is given as:
\begin{eqnarray}
L &=& 2\Bigg[\int_{0}^{R}\ \mathrm{d}r + \int_{R}^{+\infty}\ \exp\bigg(-\frac{U(r)}{k_BT}\bigg)\ \mathrm{d}x \Bigg] \\
&=& 2\Bigg[\int_{0}^{R}\ \mathrm{d}r + \int_{R}^{+\infty}\ \exp\bigg(-\frac{K(r-R)^2}{2k_BT}\bigg)\ \mathrm{d}x \Bigg] \\
&=& 2\Bigg[\int_{0}^{R}\ \mathrm{d}r + \int_{0}^{+\infty}\ \exp\bigg(-\frac{Kr^2}{2k_BT}\bigg)\ \mathrm{d}x \Bigg] \\
&=& 2\Bigg[R + \sqrt{\frac{\pi k_BT}{2K}} \Bigg].
\end{eqnarray}
The 2 outside of the brackets arises because the parameter $R$ represents the distance from the centre of the slab restraint ($x_0$).
Therefore, the effective volume of the slab is $L$ multiplied by the area of the simulation cell normal to $x$.


\subsection{2-Dimensional}
In the case of a cylindrical restraint, the 2-D distance from the axis of the cylinder is
\begin{equation}
    r(x,y) = \sqrt{(x-x_0)^2+(y-y_0)^2}
\end{equation}
with the corresponding partition function $Q$ for the 2-D flat-bottomed harmonic restraint given as;
\begin{equation}
Q = \int_{-\infty}^{+\infty} \int_{-\infty}^{+\infty}\exp\bigg(-\frac{U(r(x,y))}{k_BT}\bigg) \mathrm{d}x~\mathrm{d}y\ \\
\end{equation}
which can then be simplified into polar coordinates by taking $R(x,y)$ as $r$ and using the trigonometric relationship between $r$, $x$, $y$ and $\theta$:
\begin{equation}
Q = \int_{0}^{+\infty}\int_{0}^{2\pi}r\exp\bigg(-\frac{U(r)}{k_BT}\bigg) \mathrm{d}r~\mathrm{d}\theta . \\
\end{equation}
Therefore, the effective area of the base of the cylinder $A$ is then given by:
\begin{align}
A &=\int_{0}^{+\infty}\int_{0}^{2\pi}r\ \exp\bigg(-\frac{K(r-R)^2}{2k_BT}\bigg)\ \mathrm{d}r\ \mathrm{d}\theta\  \\
&= 2\pi\Bigg[\int_{0}^{+\infty}r\ \exp\bigg(-\frac{K(r-R)^2}{2k_BT}\bigg)\ \mathrm{d}r\Bigg] \\
&= 2\pi\Bigg[\int_{0}^{R}r\ \mathrm{d}r + \int_{R}^{+\infty}(r+R)\ \exp\bigg(-\frac{Kr^2}{2k_BT}\bigg)\ \mathrm{d}r\Bigg] \label{eq:A-2dint}\\
&= 2\pi\Bigg[\frac{1}{2}R^2 + R\sqrt{\frac{\pi k_BT}{2K}} +\frac{k_BT}{K} \Bigg]\label{eq:2d-vol}  
\end{align}
The effective volume of the cylindrical restraint can then be obtained by multiplying $A$ by the length of the simulation box in the direction parallel to the cylinder axis.
\noindent 
The effective circumference can then be obtained from the derivative of the effective area of circle with respect to the position of the restraint:
\begin{equation}
    \frac{\partial A}{\partial R} = 2\pi\Bigg[R + \sqrt{\frac{\pi k_BT}{2K}}\Bigg]
\end{equation}
which can be multiplied by the simulation cell height to give the effective surface area of the cylinder.
$R + \sqrt{\frac{\pi k_BT}{2K}}$ is the effective radius of the cylinder.

\subsection{3-Dimensional}
In the case of a spherical restraint, the 3-D distance from the centre of the sphere is
\begin{equation}
    r(x,y,z) = \sqrt{(x-x_0)^2+(y-y_0)^2+(z-z_0)^2}
\end{equation}
and the corresponding partition function $Q$ for the 3-D flat-bottomed harmonic restraint given as;
\begin{equation}
Q = \int_{-\infty}^{+\infty} \int_{-\infty}^{+\infty}\int_{-\infty}^{+\infty}\exp\bigg(-\frac{U(R(x,y,z))}{k_BT}\bigg) \mathrm{d}x~\mathrm{d}y~\mathrm{d}z\ \\
\end{equation}
which can then be simplified into polar coordinates by taking $R(x,y,z)$ as $r$ and using the trigonometric relationship between $r$, $x$, $y$, $z$, $\theta$ and $\phi$;
\begin{equation}
Q = \int_{0}^{+\infty}\int_{0}^{2\pi}\int_{0}^{\pi}r^2 sin\phi \exp\bigg(-\frac{U(r)}{k_BT}\bigg) \mathrm{d}r~\mathrm{d}\theta~\mathrm{d}\phi\ \\
\end{equation}
such that $V$ is then given by:
\begin{align}
V &= \int_{0}^{+\infty}\int_{0}^{2\pi}\int_{0}^{\pi}r^2 sin\phi \\
&\phantom{{}=11111111} \exp\bigg(-\frac{K(r-R)^2}{2k_BT}\bigg) \mathrm{d}r~\mathrm{d}\theta~\mathrm{d}\phi\ \notag\\
&= 4\pi\Bigg[\int_{0}^{+\infty}r^2\ \exp\bigg(-\frac{K(r-R)^2}{2k_BT}\bigg)\ \mathrm{d}r\Bigg] \\
&= 4\pi\Bigg[\int_{0}^{R}r^2\ \mathrm{d}r ~+ \label{eq:A-3dint} \\
&\phantom{{}=1111111}\int_{R}^{+\infty}(r+R)^2\ \exp\bigg(-\frac{Kr^2}{2k_BT}\bigg)\ \mathrm{d}r\Bigg] \notag\\
&= 4\pi\Bigg[\frac{1}{3}R^3 + \frac{\big(R^2K+k_BT\big)}{K}\sqrt{\frac{\pi k_BT}{2K}} \frac{2Rk_BT}{K} \Bigg]\label{eq:3d-vol}\\
\end{align}
\noindent 
The effective surface area of the sphere can then be computed from the derivative of the effective volume with respect to the position of the restraint:
\begin{equation}
    \frac{\partial V}{\partial R} = 4\pi\Bigg[R^2 + 2R\sqrt{\frac{\pi k_BT}{2K}} + \frac{2k_BT}{K} \Bigg]
\end{equation}





\newpage
\bibliography{manuscript}

\end{document}